\documentclass[conference]{IEEEtran}
\IEEEoverridecommandlockouts
\usepackage{cite}
\usepackage{amsmath,amssymb,amsfonts}
\usepackage{algorithmic}
\usepackage{graphicx}
\usepackage{textcomp}
\usepackage{xcolor}
\usepackage{subfig}
\usepackage{units}
\usepackage{mathtools}

\usepackage[hidelinks]{hyperref}
\usepackage{pgfplots}
\usepackage{tikz}
\usepackage[T1]{fontenc}
\usepackage[utf8]{inputenc}
\usepackage{pgfplots}
\usepackage{grffile}
\pgfplotsset{compat=newest}
\usetikzlibrary{plotmarks}
\usetikzlibrary{arrows.meta}
\usepgfplotslibrary{patchplots}

\usepackage{enumitem}

\usepackage{url}
\usepackage{adjustbox}
\usepackage[labelsep=period]{caption}
\renewcommand{\thetable}{\Roman{table}}

\usepackage{booktabs,array}
\usepackage{multirow}
\usepackage{balance}
\usepackage{colortbl}%
\newcommand{\myrowcolour}{\rowcolor[gray]{0.925}}

\definecolor{wavenet2}{RGB}{128, 0, 32}
\definecolor{wavenet}{RGB}{242, 133, 0}
\definecolor{aegan}{RGB}{87, 165, 184}
\definecolor{aegan1}{RGB}{0, 109, 91}

\newcommand\msize{3.5pt}
\newcommand\figW{4.521in}
\newcommand\figH{2.5in}
\newcommand\legfont{16pt}
\newcommand\crule[3][black,line width=2.0pt]{\textcolor{#1}{\rule{#2}{#3}}}

\begin{document}
	
	\title{Investigating Cross-Domain Losses \\for Speech Enhancement}
	
	\author{\IEEEauthorblockN{Sherif Abdulatif, Karim Armanious, Jayasankar T. Sajeev, Karim Guirguis, Bin Yang}
		\IEEEauthorblockA{University~of~Stuttgart,~Institute~of~Signal~Processing~and~System~Theory,~Stuttgart,~Germany}}
	
	\maketitle
	
	\begin{abstract}
		Recent years have seen a surge in the number of available frameworks for speech enhancement (SE) and recognition. Whether model-based or constructed via deep learning, these frameworks often rely in isolation on either time-domain signals or time-frequency (TF) representations of speech data. In this study, we investigate the advantages of each set of approaches by separately examining their impact on speech intelligibility and quality. Furthermore, we combine the fragmented benefits of time-domain and TF speech representations by introducing two new cross-domain SE frameworks. A quantitative comparative analysis against recent model-based and deep learning SE approaches is performed to illustrate the merit of the proposed frameworks.
	\end{abstract}
	
	\vspace{-0.5mm}
	\section{Introduction}
	\vspace{-0.5mm}
	\label{sec:intro}
	Speech enhancement (SE) can be defined as recovering the desired speech from various unwanted effects, such as background noise, interference and reverberation. Accordingly, SE is considered a fundamental building block in many commonplace tasks such as hearing aids, smartphones and speech recognition. However, extracting the desired speech can be challenging in realistic environments where the noise level and frequency components are similar to the desired speech.
	
	Advances in deep neural networks (DNNs) have led to leaps in performance in diverse fields such as image processing, remote sensing and medical applications \cite{ip1,ip2,rad1,rad2,rad0,rad3,rad4,med1,med2}. In an effort to overcome the intrinsic challenges in SE tasks, a large body of research similarly adopted DNNs. This has led to a surge in the number of available deep learning (DL) approaches for SE as indicated by recent studies \cite{surv1,surv2}. These approaches can be divided into two main categories. The first utilizes the original time-domain signals while the latter relies on time-frequency (TF) domain representations. 
	
	Chronologically, speech-enhancement generative adversarial networks (SEGAN) is among the first DL frameworks to adopt raw time-domain waveforms as input \cite{segan1,segan2,isegan}. This framework utilizes two networks (generator and discriminator) trained adversarially with each other. According to the original comparative results presented in \cite{segan1}, this framework surpasses the performance of traditional model-based SE approaches. Also, a non-causal adaptation of the autoregressive generative Wavenet was introduced in \cite{wavnet}. Rather than adversarial training, this architecture incorporates residual blocks together with a time-domain L1 loss. Not only did this framework improve upon the performance of the prior SEGAN, but also it is easily adaptable to variable-length input waveforms. On the whole, deeper architectures are necessary for the aforementioned approaches due to the utilized higher dimensional input space and the challenging patterns in time-domain signals.
	
	\begin{figure*}[t]
		\includegraphics[width=\textwidth]{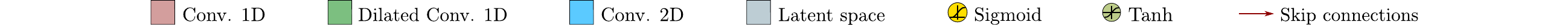}
		\subfloat[SEGAN architecure. \label{fig:segan}]{\includegraphics[width=0.297\textwidth]{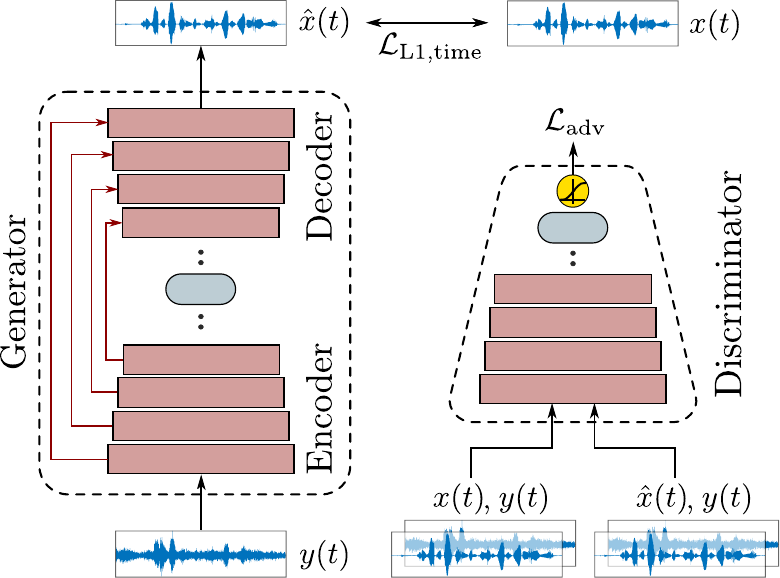}}
		\hfill\vrule height 40mm width 0.6pt\hfill
		\subfloat[Wavenet architecure. \label{fig:wavenet}]{\includegraphics[width=0.28\textwidth]{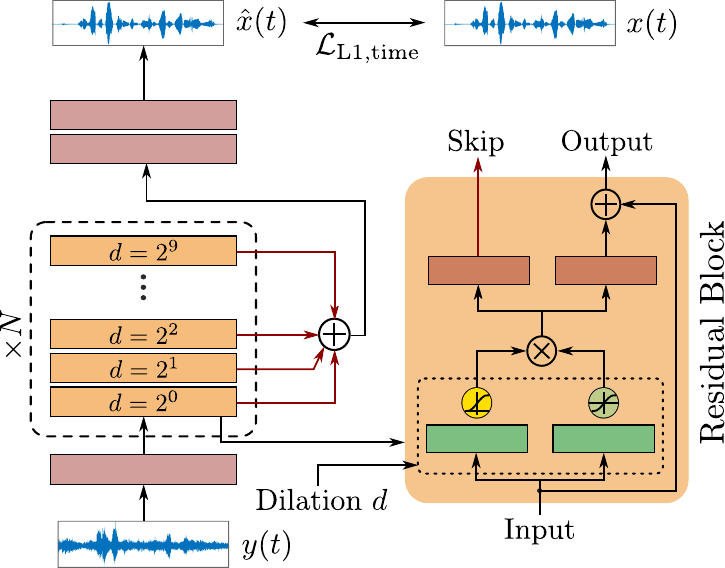}}
		\hfill\vrule height 40mm width 0.6pt\hfill
		\subfloat[AeGAN architecure. \label{fig:aegan}]{\includegraphics[width=0.39\textwidth]{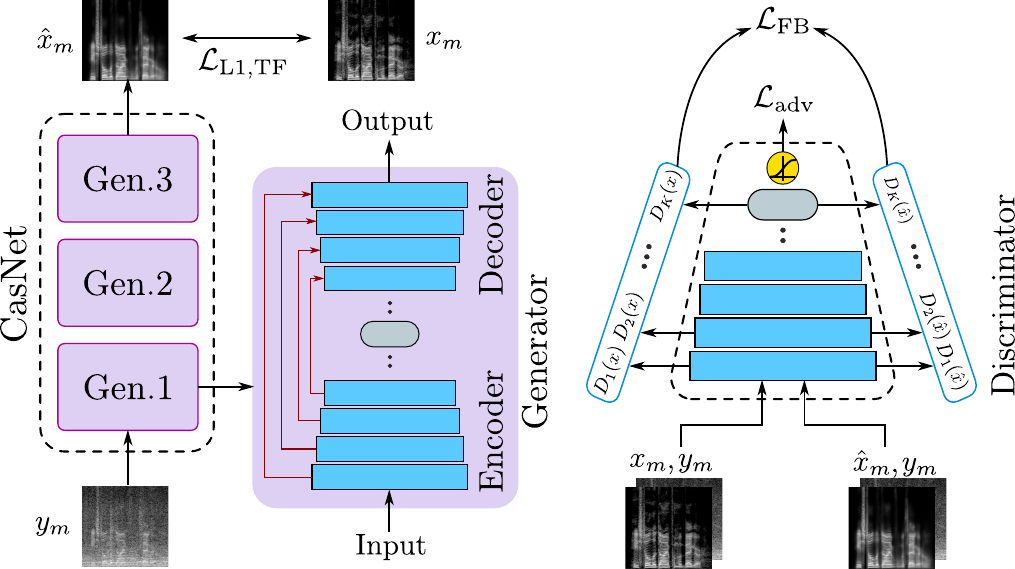}}
		\vspace{-0.5mm}\caption{Architecture details of all studied frameworks (SEGAN, Wavenet and AeGAN). The FSEGAN architecture is similar to the AeGAN with only one generator and no feature-based loss $(\mathcal{L}_{\small\textrm{FB}})$.  \label{fig:arch}}
		\vspace{-4mm}
	\end{figure*}
	
	Conversely, the second category of SE approaches convert the raw input speech data into their corresponding TF representations. This transformation serves to represent the speech data in a more perceptually oriented manner with pronounced visual features. Accordingly, the majority of model-based and DL approaches uphold this data-feeding strategy. Due to the sequential nature of speech data, usually recurrent neural networks (RNNs) are suitable for the extraction of temporal speech features \cite{lstm1,lstm2,lstm3,lstm4,lstm5}. Also, convolutional neural networks (CNNs) were studied for SE by binary masking of the corresponding TF maps \cite{cnn1,cnn2,cnn3}. Analogous to SEGAN, adversarial models were also utilized for TF-based SE. For instance, FSEGAN provided a two-dimensional adaptation of the SEGAN framework to suit TF inputs \cite{fsegan}. Building upon FSEGAN, additional adversarial approaches were introduced to provide further improvements in the resultant speech quality \cite{cganspeech1,cganspeech2,cganspeech3,aegan}. All of the aforementioned research necessitates the sole utilization of the magnitude component for SE while ignoring the phase. This will affect the quality of the reconstructed speech, i.e. how comfortable is the listening experience. However, this deterioration in the resultant speech quality does not adversely affect the speech intelligibility \cite[p.~94]{sebook}, which represents the degree of clarity of the speech content. To this end, a relatively smaller number of studies attempted dual magnitude and phase correction \cite{comp1,comp2,phasen}. Moreover, the current TF-domain approaches are confined to input tracks of fixed duration. This hinders the training and inference efficiency as longer tracks must be divided into short segments which correspond to higher computational budget.

	A new paradigm was recently proposed for SE, which explores the utilization of both time-domain and TF input components. In \cite{multidomain}, authors introduced two parallel architectures operating on the time and TF representations with a joint auxiliary decision at the end. Both architectures employ a time-domain loss operating on the final combined output. Few studies attempted penalizing a time-domain framework with a TF-domain loss \cite{segan2,timejoint,poco}. Moreover, some recent frameworks utilized a binary masking operation to enhance input TF representations which are then transformed back to the time-domain \cite{joint1,joint}. Similarly, a single loss function is used to penalize the whole framework in the time-domain. Thus, despite utilizing both domains as input components, the above frameworks are trained solely using a single-domain loss function. To the best of our knowledge, no explicit investigation has been conducted regarding the impact of simultaneous utilization of cross-domain loss functions for SE.
	
	In this work, we attempt to combine the fragmented benefits of time and TF-domain SE frameworks. To this end, we empirically investigate the impact of single-domain loss functions and their cross-domain counterparts. Namely, we incorporate the time-domain loss component into SE frameworks operating on TF representations and vice versa. Further, we separately examine the influence of each loss component on the resultant speech quality and speech intelligibility. To validate the advantages of the proposed cross-domain models, a quantitative comparative analysis between recent SE frameworks and the proposed cross-domain approaches is presented over different signal-to-noise-ratios (SNRs).
	
	\vspace{-1mm}
	\section{Method\label{sec:method}}
	\vspace{-0.5mm}	
	In this section, the SE approaches investigated in the conducted comparative analysis will be briefly introduced. An overview of these frameworks is represented in Fig.~\ref{fig:arch}. Moreover, the proposed cross-domain loss functions will be subsequently presented in details.
	\vspace{-0.5mm}
	\subsection{Time-domain Approaches}
	\textbf{SEGAN} \cite{segan1} is a conditional adversarial framework composed of two networks, the generator $G$ and the discriminator $D$, as depicted in Fig.~\ref{fig:segan}. The generator is a one-dimensional 22-layer U-net architecture which receives as input a fixed-length corrupted time-domain speech signal $y(t)$. It is tasked with transforming it into a corresponding enhanced output $\hat{x}(t)=G\left(y(t)\right)$. The discriminator network, with identical architecture as $G$'s encoder, receives either the ground-truth clean speech signal $x(t)$ or the generated speech output $\hat{x}(t)$ together with the corrupted input $y(t)$. It acts as a binary classifier as it attempts to distinguish which input-pair is real and which is fake. This motivates $G$ to improve its own performance by producing more realistic output which fools $D$, while conversely $D$ attempts to improve its classification performance. This adversarial training component is represented by the following loss function:
	\begin{equation}
		\vspace{-1mm}
		\begin{split}
			\min_{G} \max_{D} \mathcal{L}_{\small\textrm{adv}} = \min_{G} \max_{D} & \; \mathbb{E}_{x(t),y(t)} \left[\textrm{log} D(x(t),y(t)) \right] + \\
			& \;\mathbb{E}_{\hat{x}(t),y(t)} \left[\textrm{log} \left( 1 - D\left(\hat{x}(t),y(t)\right) \right) \right]
		\end{split}
		\label{eq:gan}
		\vspace{-1mm}
	\end{equation}
	while an additional time-domain loss component is utilized to further improve the generator training:
	\begin{equation}
		\vspace{-0.5mm}
		\mathcal{L}_{\small\textrm{L1,time}} = \mathbb{E}_{x(t),\hat{x}(t)} \left[\lVert{x(t) - \hat{x}(t)}\rVert_1\right]
		\label{eq:L1--time}
		\vspace{-0.5mm}
	\end{equation}
	\begin{figure*}[t]
		\subfloat[CD-AeGAN architecture and losses.\label{fig:cd_aegan}]{\includegraphics[width=0.62\textwidth]{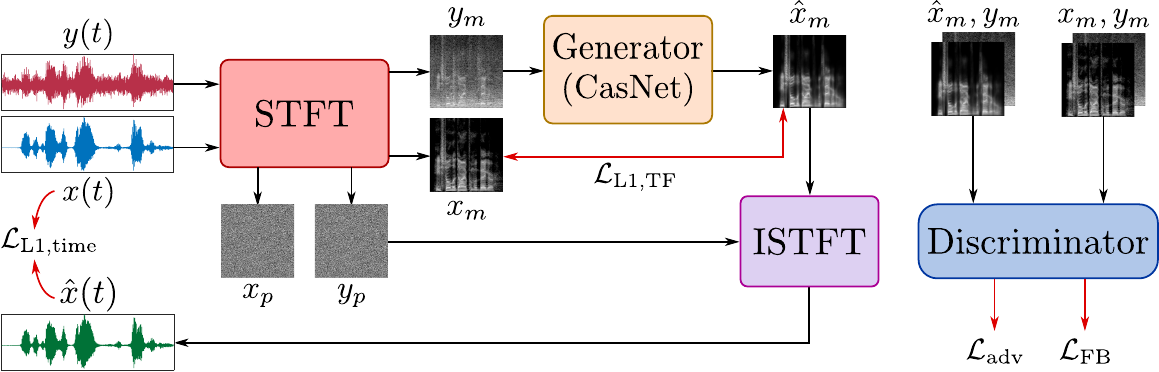}}
		\hfill \vrule height 34mm width 0.6pt \hfill
		\subfloat[CD-Wavenet architecture and losses.\label{fig:cd_wavenet}]{\includegraphics[width=0.35\textwidth]{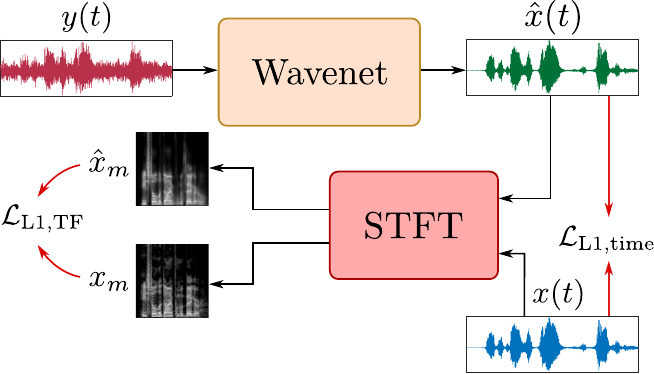}}
		\caption{SE cross-domain architectures with their respective losses. The weights given to cross-domain losses are chosen to reflect equal importance and thus penalization in both domains. All STFT blocks utilize dynamic time resolution technique \cite{aegan}, to embed all training tracks into 256 $\times$ 256 TF representations. \label{fig:cd_models}}
		\vspace{-4mm}
	\end{figure*}
	\textbf{Wavenet} \cite{wavnet} is an audio domain adaptation of the autoregressive "PixelCNN" generative model for natural images \cite{pixelcnn}. It is composed of consecutive residual blocks preceded and followed by a series of one-dimensional convolutional layers. The residual blocks utilize dilated convolutional layers with exponentially increasing dilation factors resulting in a growing receptive field. Due to the sequential nature of time-series speech data, it is beneficial to capture long-term dependencies in order to improve the resultant speech quality. For this purpose, Wavenet employs gated activation units within the residual blocks, as illustrated in Fig.~\ref{fig:wavenet}. Finally, the resultant enhanced speech signal is penalized by an L1 loss, identical to that in Eq.~\ref{eq:L1--time}. A particular advantage of this approach is the capability of accommodating variable-length inputs, whereas SEGANs are restricted to fixed-length data as they employ an encoder-decoder generator architecture.
	
	\vspace{-0.5mm}
	\subsection{Time-Frequency Approaches}
	\textbf{FSEGAN} \cite{fsegan} is a two-dimensional adaptation of the SEGAN framework. It follows the same design principles of conditional adversarial networks with a generator $G$ (16-layered U-net architecture) and a discriminator $D$ trained simultaneously in competition with each other. In this approach, the inputs to the generator are TF-magnitude representations calculated by applying a short-time Fourier transform (STFT) on the raw speech data. Compared to prior time-domain approaches, the advantages of this framework are twofold. First, the TF-magnitude component is a visual interpretation of human speech phonetics. This assists in enhancing the performance of DL-related SE approaches, as reported in \cite{fsegan}. Further, this TF embedding reduces the required computational resources due to the efficient two-dimensional convolutional layers. For training this network, a similar adversarial loss function to that defined in Eq.~\ref{eq:gan} is utilized albeit with the TF-magnitude representations $y_m$, $x_m$ and $\hat{x}_m$ representing the corrupted inputs, clean targets and predicted outputs, respectively. To further enhance the quality of the resultant TF representations, an additional L1 loss is used to penalize the pixel-wise discrepancies between the outputs $\hat{x}_m$ and targets $x_m$ as expressed by: 
	\begin{equation}
		\mathcal{L}_{\small\textrm{L1,TF}} = \mathbb{E}_{x_m,\hat{x}_m} \left[\lVert{x_m - \hat{x}_m}\rVert_1\right]
		\label{eq:L1-TF}
	\end{equation}
	
	\textbf{AeGAN} \cite{aegan} is an improved architecture which extends FSEGAN by introducing an additional non-adversarial feature-based loss function together with an enhanced generator consisting of an end-to-end concatenation of three U-nets (CasNet), as shown in Fig.~\ref{fig:aegan}. Moreover, the feature-based loss function $(\mathcal{L}_{\small\textrm{FB}})$ exploits the visual patterns in the input TF representations to produce globally consistent results. This is achieved by utilizing the discriminator as a trainable feature-extractor network and penalizing the discrepancies between the extracted feature-maps of the outputs $\hat{x}_ m$ and their ground-truth counterparts $x_m$. This loss is defined as:
	\begin{equation}
		\mathcal{L}_{\small\textrm{FB}} = \sum_{k = 1}^{K}  \lambda_{k} \lVert{D_k\left(x_m\right) - D_k\left(\hat{x}_m\right)}\rVert_1
	\end{equation}
	with $D_k$ representing the extracted feature-map from the $k^{th}$ layer of the discriminator. $K$ and $\lambda_k$ are the number of layers and the individual weights given to each layer, respectively.
	\vspace{-1mm}
	\subsection{Cross-domain Loss Functions}
	In this study, we attempt to combine the benefits of both time-domain and TF-based SE approaches. To achieve this, we utilize from the prior two families of approaches the Wavenet and AeGAN as baselines and expand them with additional cross-domain loss functions, as illustrated in Fig.~\ref{fig:cd_models}. Below are the two resultant frameworks:
	
	\begin{itemize}[label=$\bullet$,wide = 0pt]
		\setlength{\itemindent}{0.5em}
		\setlength{\itemsep}{1.5pt}
		\setlength{\parskip}{1pt}
		\item \textbf{Time $\mathbb{\rightarrow}$ TF loss}: The first framework, referred to as cross-domain Wavenet (CD-Wavenet), is an extension of the original Wavenet architecture. It implements an STFT on both the enhanced and target speech signals. The resultant TF-magnitude components are penalized using the cross-domain loss function $\mathcal{L}_{\small\textrm{L1,TF}}$, presented previously in Eq.~\ref{eq:L1-TF}.  Since the TF-magnitude is a direct representation of speech phonetics, we hypothesize that additionally penalizing signals in the TF-domain will help to preserve the perceptual speech features, thus, enhancing the speech intelligibility with no adverse effect on the resultant quality.
		\item \textbf{TF $\mathbb{\rightarrow}$ Time loss}: Cross-domain AeGAN (CD-AeGAN) is the second investigated framework. It reverts back to the original time-domain waveforms via an inverse short-time Fourier transform (ISTFT) operation applied to the output and target TF-magnitude representations, together with the input noisy phase $y_p$. Discrepancies in the resultant time-series are then penalized using the aforementioned $\mathcal{L}_{\small\textrm{L1,time}}$ loss in Eq.~\ref{eq:L1--time}. This guides the generator network to produce speech tracks that not only matches the perceptual features in the TF-domain, but also the audible aspects of the original time-domain resulting in enhanced speech quality. The claim that time-domain loss inherits phase information, and thus directly impacts the speech quality, whereas speech intelligibility is rather affected by TF-magnitude is consistent with previous studies \cite{qi_study,surv}. 
	\end{itemize}
	
	\section{Datasets and Experiments}
	\label{sec:exp}
	
	To investigate the impact of cross-domain loss functions in SE, we conduct a comparative study using the Voice Bank corpus dataset \cite{Yamagishi2019CSTRVC} as the ground-truth reference tracks. Additive noise from the DEMAND and QUT-TIMIT datasets \cite{demand,qutnoise} were used to create the corrupted speech input tracks according to the following signal-to-noise ratios (SNRs): 0 and 5 dB. For training the different networks, data from 30 speakers (175 sentences) corrupted with 12 noise conditions were utilized (total of 49,510 tracks). As for the testing dataset, it consists of 20 new speakers reading 50 new sentences with 7 different noise conditions (1,000 tracks per SNR value). All input speech tracks are sampled at 16 kHz.
	
	We quantitatively examine the performance of the above-described DL approaches together with two model-based approaches: the Wiener filter \cite{wiener} as a baseline and a recent Bayesian MMSE technique introduced in 2018 \cite{bayesian_2018}. The SEGAN framework was trained on fixed one-second duration tracks, whereas the Wavenet utilized variable-length tracks. The TF-domain approaches were extended to incorporate the dynamic time resolution technique presented in \cite{aegan}. This serves to embed tracks of variable duration into TF-magnitude embeddings of a fixed $256 \times 256$ dimensionality. To ensure a fair comparison, all trainable models were trained for 50 epochs using the architectures and hyper-parameter settings recommended in their respective original publications. 
	
	To investigate both the resultant speech quality and intelligibility, multiple metrics were used in the comparative analysis. For speech quality assessment, the segmental SNR (SSNR) \cite{ssnr}, the perceptual evaluation of speech quality (PESQ) \cite{pesq}, the mean opinion score (MOS) predictions of the signal distortion (CSIG), background noise (CBAK) and the overall condition (COVL) \cite{mos} are utilized. With regards to human speech intelligibility, we evaluate the word error rate (WER) of a pre-trained Deep Speech model \cite{deepspeech} as well as the short-time objective intelligibility measure (STOI) \cite{stoi}.

	\begin{table}[!t]
		\vspace{-1mm}
		\caption{Quantitative comparison of different approaches on the test set. Bold font indicates the best scores. The metric (1-WER) is utilized to represent improvements in the resultant intelligibility score (higher is better). \label{tab:test}}
		\centering
		\setlength\arrayrulewidth{1.5pt}
		\fontsize{23}{23}
		\selectfont
		\bgroup
		\def\arraystretch{1.5}
		\resizebox{\columnwidth}{!}{%
			\begin{tabular}{r|ccccccc}
				\hline
				\multirow{3}{*}{Model} & \multicolumn{7}{c}{(a) SNR 0 dB}\\
				&PESQ & CSIG & CBAK & COVL & SSNR & STOI & 1-WER\\
				&   &   &   &   & [dB] & [\%] & [\%]\\
				\hline
				\myrowcolour Wiener & 1.83 & 1.35 & 2.04 & 1.49 & -0.52 & 63.1 & 12.2\\
				Bayesian & 2.05 & 1.84& 1.86 & 1.75 & -0.37 & 64.2 & 20.4\\
				\hline
				\myrowcolour SEGAN & 2.16 & 2.47 & 2.18 & 2.20 & 0.24 & 67.8 & 28.5\\
				Wavenet & 2.34 & 3.33 & 2.59 & 2.78 & 2.44 & 69.2 & 51.2\\
				\myrowcolour CD-Wavenet & 2.43 & 2.98 & 2.62 & 2.64 & 2.74 & 69.7 & \textbf{58.6}\\
				\hline
				FSEGAN & 2.33 & 3.26 & 2.50 & 2.75 & 0.59 & 68.2 & 46.9\\
				\myrowcolour AeGAN & 2.48 & 3.57 & 2.68 & 2.99 & 1.59 & 69.9 & 54.2\\
				CD-AeGAN & \textbf{2.53} & \textbf{3.61} & \textbf{2.70} & \textbf{3.06} & \textbf{3.03} & \textbf{70.2} & 56.6\\
				\hline\hline
				
				\multirow{3}{*}{Model} & \multicolumn{7}{c}{(b) SNR 5 dB}\\
				&PESQ & CSIG & CBAK & COVL & SSNR & STOI & 1-WER\\
				&   &   &   &   & [dB] & [\%] & [\%]\\
				\hline
				\myrowcolour Wiener & 2.20 & 1.85 & 2.38 & 1.95 & 1.23 & 72.2 & 35.9\\
				Bayesian & 2.38 & 2.32 & 2.23 & 2.19 & 1.82 & 73.1 & 37.6\\
				\hline
				\myrowcolour SEGAN & 2.50 & 2.83 & 2.52 & 2.56 & 2.05 & 74.5 & 47.5\\
				Wavenet & 2.73 & 3.74 & 2.91 & 3.19 & 4.03 & 75.9& 62.8\\
				\myrowcolour CD-Wavenet & 2.77 & 3.49 & 2.96 & 3.09 & 4.73 & 76.7 & \textbf{70.9}\\
				\hline
				FSEGAN & 2.68 & 3.68 & 2.83 & 3.16 & 2.32 & 75.7 & 64.8\\
				\myrowcolour AeGAN & 2.87 & 3.91 & 3.03 & 3.37 & 3.64 & 79.2 & 68.9\\
				CD-AeGAN & \textbf{2.92} & \textbf{3.93} & \textbf{3.10} & \textbf{3.40} & \textbf{4.87} & \textbf{79.6} & 70.1\\
				\hline
			\end{tabular}
		}
		\egroup
		\vspace{-4mm}
	\end{table}
	\section{Results and Discussion}
	\label{sec:results}
	
\begin{figure}[!t]
		\vspace{-2.5mm}
		\centering
		\resizebox{\columnwidth}{!}{\input{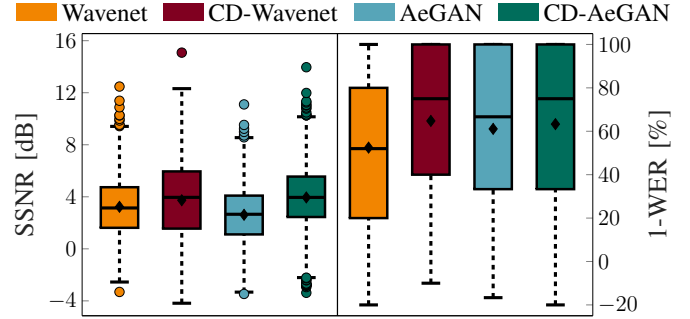}}
		\caption{Boxplots of cross-domain models in comparison to their original counterparts over both SNRs (0, 5 dB). The SSNR (left) represents the speech quality. The WER (right) represents the speech intelligibility. The quantity (1-WER) can be negative as in these cases, the accumulated errors exceed the number of words in the reference sentence. \label{fig:boxplots}}
		\vspace{-4mm}
	\end{figure}

	Table~\ref{tab:test} and Fig.~\ref{fig:boxplots} present the quantitative evaluations of the proposed cross-domain SE frameworks in comparison to approaches utilizing single-domain losses. Additionally, comparisons were carried out against other model-based approaches and DL-based time-domain and TF-domain approaches. The recent Bayesian MMSE model-based approach achieves comparable performance to the SEGAN framework with respect to both the speech quality and intelligibility metrics. The AeGAN framework achieves a significant enhancement of the resultant speech tracks compared to the FSEGAN framework as reflected across all metric scores. This is attributed to the incorporation of the cascaded generator network architecture as well as the feature-based loss function. For instance, at SNR 0 dB improvements from 46.9\% to 54.2\% and from 2.33 to 2.48 were observed for the WER and PESQ, respectively. With regards to time-domain SE approaches, the Wavenet framework introduces the capability of enhancing speech tracks of variable-lengths. Despite of this, it outperforms the prior SEGAN framework which deals with fixed-length input tracks. 
	
	To investigate the impact of cross-domain loss functions in SE, the highest performing networks from both the time and TF-based family of DL techniques were singled out. Namely, the AeGAN and Wavenet frameworks. These approaches were then expanded upon by cross-domain loss functions and compared to their unaltered counterparts. In CD-AeGAN, the output TF-magnitude representations from AeGAN are transformed back to the time-domain before being additionally penalized by an L1 loss function. This translates into an improvement of the SSNR score compared to AeGAN by approximately 1.44 and 1.23 dB for SNR 0 and 5 dB, respectively. Also, an enhancement of the PESQ score is observed with a slight improvement in the CSIG, CBAK and COVL. Furthermore, a marginal or no improvement in the speech intelligibility metrics, represented by WER and STOI, is perceived. The second investigated cross-domain approach expands upon the conventional Wavenet framework with an additional pixel-wise loss in the TF-magnitude. In contrast to CD-AeGAN, penalizing the TF-magnitude representations noticeably enhances the speech intelligibility of the resultant tracks. Specifically, the WER was improved by approximately 8\% for both SNR values. Conversely, marginal improvement can be observed to the speech quality metrics. On the whole, the CD-AeGAN outperforms the CD-Wavenet in most of the metrics except for the WER where the CD-Wavenet achieves the best score. This agrees with recent literature hypothesizing that penalizing the time domain after the enhancement of TF representations is beneficial to the overall quality metrics \cite{joint}. 
	
	The above results indicate that time-domain and TF-based loss components have distinct strengths.  Specifically, penalizing the speech data in the TF-domain positively influences the resultant speech intelligibility, whereas time-domain loss functions directly impact the speech quality. From another perspective, time-domain approaches benefit from implicitly enhancing the speech phase component which positively impacts the quality metrics, whereas TF-based approaches exploit the perceptual features in the magnitude representations, thus, enhancing the ineligibility of the speech tracks. Consequently, cross-domain SE approaches would assist in bridging the gap between the distinct advantages of single-domain frameworks. 
	\vspace{-2mm}
	\section{Conclusion}
	\label{sec:conc}
	In this work, we present an investigative study regarding the impact of simultaneously enhancing speech signals in both time and TF domains in comparison to single-domain approaches. To achieve this, we expand the conventional Wavenet and AeGAN frameworks with corresponding cross-domain loss functions. Quantitative evaluations have illustrated that penalizing time-domain losses directly influence the resultant speech quality. In contrast, TF-based loss functions impact the intelligibility of the output speech. In the future, this may lead to the development of tailored architectures to resolve the drawbacks of single-domain frameworks. For instance, this could combine the benefit of time-domain approaches with respect to implicitly enhancing the phase component while concurrently exploiting the perceptual information in the TF-magnitude representations.  
	
	\bibliographystyle{IEEEbib}
	
\end{document}